\documentclass[12pt]{article}

\usepackage{amsmath}
\usepackage{amssymb}

\newtheorem{theorem}{Theorem}[section]

\newtheorem{prop}{Proposition}[section]
\newtheorem{definition}{Definition}[section]
\numberwithin{equation}{section}

\begin{document}

\begin{center}
 Paragrassmann Algebras as
Quantum Spaces
\\
Part I: Reproducing Kernels
 \vskip 0.5cm
 Stephen Bruce Sontz
  \\
 Centro de Investigaci\'on en Matem\'aticas, A.C.
 \\
 CIMAT
  \\
 Guanajuato, Mexico
  \\
 email: sontz@cimat.mx
 \end{center}

\begin{abstract}
\noindent
Paragrassmann algebras are given a sesquilinear form for
which one subalgebra becomes a Hilbert space known as the
Segal-Bargmann space.
This Hilbert space as well as the ambient space of the
paragrassmann algebra itself are shown to have reproducing kernels.
These algebras are not isomorphic to algebras of functions
so some care must be taken in defining what ``evaluation at a
point'' corresponds to in this context.
The reproducing kernel in the Segal-Bargmann space is shown
to have most, though not all, of the standard properties.
These quantum spaces provide non-trivial examples of spaces
which have a reproducing kernel but which are not spaces of functions.
\end{abstract}


\section{Introduction}

This paper is inspired in large measure by the work in \cite{csq} on paragrassmann algebras.
We begin in Sections~1-4 by reviewing some of the material in \cite{csq}, though sometimes
re-working that presentation by using our own notation and sometimes by
making mild generalizations.
We also prove some basic propositions for later use and
explain in detail the conjugation we will be using.
This makes the paper more self-contained logically.
However, see \cite{csq} for references to previous works on this topic in mathematics
and physics.
We note that the deformation parameter $q$ in this paper is non-zero and complex, while in \cite{csq}
it lies on the unit circle in the complex plane.
But more importantly, the conjugation used here is different from that in \cite{csq}.
So strictly speaking this paper treats topics not discussed in \cite{csq}, though there are
ideas in common.
The core material of the paper starts in Section~5 where reproducing kernels are defined
and discussed in the context of a Segal-Bargmann space that we define as a subalgebra
of a paragrassmann algebra.
The Segal-Bargmann (or coherent state) transform is introduced in Section~6,
and its relation to the reproducing kernel is proved.
We follow in Section~7 with a proof of the existence of the reproducing kernel in the full space of
paragrassmann variables.
This might seem to be a rather surprising result since this is a non-commutative algebra in general unlike the
Segal-Bargmann space, which is a commutative algebra though also not isomorphic to an algebra of functions.
However, reproducing kernels in the finite dimensional case are quite common and,
as we shall see, this is even the generic case in some sense to be specified later.
In the last section we discuss some possible avenues for future research.
One of these possibilities, the definition and study of Toeplitz operators in this context, will be the topic of
a forthcoming paper \cite{part2}.

\section{Preliminaries}

Throughout this article 
we take $l$ to be an integer with $l \ge 2$.
We put $q_l = e^{2 \pi i / l}$, a primitive $l$-th root of unity in the complex plane $\mathbb{C}$.
(N.B. Our parameter $l$ corresponds to $k^\prime$ in \cite{csq}.)
We take the set $\{ \theta, \overline{\theta} \}$ of two elements and consider
the free algebra over the field of complex numbers $\mathbb{C}$ generated 
by this set. It is denoted by $\mathbb{C}  \{ \theta, \overline{\theta} \} $.
It is also called the algebra of complex polynomials in the two non-commuting variables
$ \theta, \overline{\theta} $ which satisfy no relation whatsoever.
In this paper all spaces are vector spaces over the field $\mathbb{C}$,
and all algebras are \textit{unital}, that is, have an identity element.
Moreover, algebra morphisms map the identity element in the domain to
the  identity element in the codomain.

As in \cite{csq}
we define the \textit{paragrassmann algebra associated to $l$} to be the quotient algebra
\begin{equation}
\label{define_para}
PG_l = PG_l(\theta, \overline{\theta} ) := \mathbb{C}  \{ \theta, \overline{\theta} \} / \langle \theta^l,
\overline{\theta}{}^l, \theta \overline{\theta} - q_l \overline{\theta} \theta \rangle.
\end{equation}
Here as usual the notation $\langle \, \cdot \, \rangle$ means the two-sided ideal generated by the
elements listed inside the braces.
We let $ \theta, \overline{\theta} $ also denote the quotients (i.e., equivalence classes) of these two
elements in $PG_l$.
Seen this way  $ \theta$ and $\overline{\theta} $ are nilpotent elements in $PG_l$, each having 
order of nilpotency $l$.
They do not commute, since  $ \theta \overline{\theta} = q_l \overline{\theta} \theta$ in $PG_l$ and $q_l \ne 1$.
(The case $l=1$ has been excluded not so much because $q_l =1$ in that case, but really because
$ \theta = \overline{\theta} =0 $ in that case and so $PG_1 = \mathbb{C}$, a trivial case we wish to
exclude.)
However, in the case $l=2$ we have $q_2 = -1$ and so $ \theta$ and $\overline{\theta} $ anti-commute.
 Also, $ \theta^2 = \overline{\theta}{}^2 =0$.
So the case $l=2$ corresponds to two \textit{grassmann variables} in the standard definition of this term.
In \cite{csq} the paragrassmann algebra studied is $PG_l$. Notice that in this algebra
the order of nilpotency of the variables $\theta$ and $\overline{\theta}$ determines the parameter
$q_l$ in the commutation relation, and conversely.
However, we would like to generalize slightly this concept in the following definition.

\begin{definition}
Let $l \ge 2$ be an integer and $q \in \mathbb{C}$.
The {\rm paragrassmann algebra} $PG_{l,q}$ with {\rm paragrassmann variables} $\theta$ and $\overline{\theta}$
is defined by
\begin{equation}
\label{define_para_lq}
PG_{l,q} =
PG_{l,q}(\theta, \overline{\theta})  := \mathbb{C}  \{ \theta, \overline{\theta} \} / \langle \theta^l, \overline{\theta}{}^l, \theta \overline{\theta}
- q \overline{\theta} \theta \rangle
\end{equation}
\end{definition}
Clearly, $PG_l$ is the special case of $PG_{l,q} $ when $q = q_l$.
We will be studying $PG_{l,q} $.
The equation $\theta \overline{\theta} - q \overline{\theta} \theta = 0$ in $PG_{l,q}$
is called the \textit{$q$-commutation relation}, while $\theta^l =0$ and $ \overline{\theta}{}^l =0$ in $PG_{l,q}$
are called the \textit{nilpotency conditions}.
We note that 
\begin{equation}
\label{a-basic-equality}
PG_{l,q}(\theta, \overline{\theta}) = PG_{l,q^{-1}}(\overline{\theta}, \theta) 
\end{equation}
for $ q \ne 0$.
This is an \textit{equality} (not just an isomorphism) of sets, of vector spaces and also of algebras.
Notice that the order of the paragrassmann variables is different on the two sides of this equality.
Since neither element in the pair of paragrassmann variables $\theta, \overline{\theta}$ is more fundamental than the other (each
being the conjugate of the other in a conjugation defined later)
we also have an \textit{isomorphism}
\begin{equation}
\label{a-basic-isomorphism}
    PG_{l,q}(\theta, \overline{\theta}) \cong PG_{l,q^{-1}}(\theta, \overline{\theta})
\end{equation}
of algebras, where
the isomorphism on the generators maps $\theta \mapsto \overline{\theta}$ and  $\overline{\theta} \mapsto \theta$.
This is then extended multiplicatively to the basis $AW$, defined below, and then \textit{linearly} to $PG_{l,q}(\theta, \overline{\theta})$.
Notice that the order of the paragrassmann variables is the same on the two sides of this isomorphism.
One can combine the equality (\ref{a-basic-equality}) and the isomorphism
(\ref{a-basic-isomorphism}) to get other identifications.
The moral of these basic facts is that at this stage one can not distinguish between the 
creation and annihilation elements, where one of these should be 
$\theta$ while the other should be $\overline{\theta}$.
It is only after quantization that such a distinction can be made by examining the quantizations of 
$\theta$ and $\overline{\theta}$.
We will see this in detail in \cite{part2} in the quantization given by the Toeplitz operators.
So the algebra $PG_{l,q} $ could be viewed as `classical' object 
in some sense even though it is also a `quantum' object, that is, it is not commutative.

The case $q=0$ is different as would be expected.
In any quantization scheme (and there are many) this `classical' algebra for $q=0$, namely $PG_{l,0} $, gives rise to what
could be called a `quantum' free probability theory of paragrassmann variables.

We will be using the following index set throughout:
$$
        I_l = \{0, 1, \dots, l-1 \}.
$$
When an index, say $i$, is given without an explicit index set, we assume  $i \in I_l$.

The \textit{Segal-Bargmann (or holomorphic) space} is defined to be
\begin{equation}
\label{def_sb}
\mathcal{B}_H = \mathcal{B}_H(\theta) := \mathrm{span}_{\mathbb{C}} \, \{ \theta^i \, \vert \, i \in I_l \}.
\end{equation}
Similarly, the \textit{anti-Segal-Bargmann (or anti-holomorphic) space} is defined to be
\begin{equation}
\label{def_anti_sb}
\mathcal{B}_{AH} =\mathcal{B}_{AH}(\overline{\theta}) := \mathrm{span}_{\mathbb{C}} \, \{ \overline{\theta}{}^i \, \vert \, i \in I_l \}.
\end{equation} 

The Segal-Bargmann space is not only a vector subspace of $PG_{l,q}(\theta, \overline{\theta})$; it is also
a subalgebra. 
Actually, it is a commutative subalgebra isomorphic to the truncated polynomial
algebra $\mathbb{C} [\theta] \, / \langle \theta^l \rangle$.
Similarly, the anti-Segal-Bargmann space is a commutative subalgebra isomorphic to the exact same truncated polynomial
algebra, although this is usually written as $\mathbb{C} [\overline{\theta}] \, / \langle \overline{\theta}{}^l \rangle$.

The subspace $\mathcal{S} = \mathcal{B}_H + \mathcal{B}_{AH}$ of $PG_{l,q} $ plays a special role.
Note that this is not a subalgebra.
Also this is not a direct sum but nearly so, since
$\mathcal{B}_H \cap \mathcal{B}_{AH} = \mathbb{C}1$.
We define a conjugation operation in $\mathcal{S}$ by $(\theta^i)^* := \overline{\theta}{}^i$ and
$(\overline{\theta}{}^i)^* := \theta^i$ and extend \textit{anti-linearly} to all of  $\mathcal{S}$.
For the time being we do not introduce a conjugation operation in $PG_{l,q}$ though we will do
this later on by extending this rather natural conjugation in $\mathcal{S}$ to $PG_{l,q}$.

We have two canonical bases of $PG_{l,q}$ provided that $q \ne 0$.
First, there is the \textit{Wick basis}
$$
    W = \{ \overline{\theta}{}^i \theta^j \, \vert \, i,j \in I_l \},
$$
which is also a basis when $q=0$.
Second, there is the \textit{anti-Wick basis}
$$
    AW = \{ \theta^i \overline{\theta}{}^j \, \vert \, i,j \in I_l \}
$$
for the case $q \ne 0$.
Of course, we are using here the usual convention that $\theta^0 =1$
and $\overline{\theta}{}^0 = 1$.
Notice that the elements of these two bases are not the same, except when $q=1$.
For $q \ne 1$ their only common element is the identity element $1$.
Here we follow \cite{csq} by saying that an expression with all factors of 
$\theta$ to the right (respectively, left) of all factors
of $\overline{\theta} $ is in \textit{Wick} (respectively, \textit{anti-Wick}) order.
In physics (e.g., see \cite{glimm-jaffe}) 
the original definition is that an expression with all annihilation operators
to the right of all creation operators is in Wick order.
Since we have no way for identifying at this level which variable corresponds to annihilation,
our present definition is a rather arbitrary choice whose only virtue is that it agrees with \cite{csq}.
Clearly, we have $\mathrm{dim}_{\mathbb{C}} \, PG_{l,q} = l^2$.

In the rest of this article, we will consider the case $q \in \mathbb{C} \setminus \{ 0 \}$.
The results in this paper do not depend directly on the specific value of the parameter $q$.
It seems that the principle role of the $q$-commutation relation
(in conjunction with the nilpotency conditions) is to force the space under consideration
to have finite dimension $l^2$ with the very specific standard basis $AW$.

The integral is a linear functional $PG_{l,q} \to \mathbb{C}$ defined
on the basis $AW$ by
$$
 \int \! \! \! \int  d \theta \, \, \theta^i \overline{\theta}{}^j  \, \, d \overline{\theta} := \delta_{i,l-1}  \delta_{j,l-1},
$$
where $\delta_{ab}$ is the Kronecker delta of the integers $a,b$.
This is a Berezin type integral, by which is meant that only the highest non-zero power element
$\theta^{l-1} \overline{\theta}{}^{l-1}$ has a non-zero integral.

\section{Conjugation}

We next introduce a conjugation (or $*$-operation) in $PG_{l,q}$ by expanding an arbitrary $f \in PG_{l,q}$ in the
basis $AW$ as
\begin{equation}
    f = \sum_{i,j} f_{ij} \theta^i \overline{\theta}{}^j,
\end{equation}
where the coefficients $f_{ij} \in \mathbb{C}$ are uniquely determined.
Then we define the \textit{conjugation} of $f$ by
\begin{equation}
\label{define-star}
f^* :=  \sum_{i,j} f_{ij}^* \theta^j \overline{\theta}{}^i.
\end{equation}
(The usual complex conjugate of $\lambda \in \mathbb{C}$ is denoted by $\lambda^*$.)
This gives the expansion of $f^*$ in the same basis $AW$.
This is an anti-linear operation.
It also immediately follows that $f^{**} = f$, that is, this operation
is an involution.
Note that  the conjugation 
being an involution depends on $\theta$ and $\overline{\theta}$
having the same order of nilpotency.
Actually, as another immediate consequence of the definition (\ref{define-star}) we also have that
$( \theta^i  \overline{\theta}{}^j )^*  =  \theta^j \overline{\theta}{}^i $ and in particular the relations
\begin{equation}
\label{more-relations}
(\theta^i)^* = \overline{\theta}{}^i \quad \mathrm{and} \quad (\overline{\theta}{}^j)^* = \theta^j \! 
\end{equation}
as promised earlier in Section~2.
The action of the conjugation on elements in the basis $W$ is given by
\begin{equation}
\label{star-on-W-basis}
      ( \overline{\theta}{}^i   \theta^j )^* = ( q^{-i j} \,  \theta^j  \overline{\theta}{}^i )^* = 
     ( q^{-i j} )^*  \theta^i \overline{\theta}{}^j
      = (q^{-ij})^* q^{  i j}  \overline{\theta}{}^j \, \theta^i 
       = \left( \dfrac{q}{q^*} \right)^{ij} \overline{\theta}{}^j \theta^i .
\end{equation}
So for $q \in \mathbb{R} \setminus \{ 0 \}$ the two bases $W$ and $AW$ are treated in a similar manner by the conjugation.
But the expression on the far right of (\ref{star-on-W-basis}) has no limit when $ q \to 0$ and $q$ is complex.

We note that the subspace $\mathcal{S}$ is invariant under the conjugation and that this operation interchanges
its subalgebras $\mathcal{B}_H$ and $\mathcal{B}_{AH}$.
As noted earlier,  these two subalgebras are isomorphic as algebras under the $\mathbb{C}$-linear map 
 $\mathcal{B}_H \to \mathcal{B}_{AH}$  induced by $ \theta^i \mapsto \overline{\theta}{}^i$ for all $i \in I_l$.
Moreover,  the conjugation gives us an explicit anti-isomorphism
 (and its inverse) between these subalgebras, because of the relations (\ref{more-relations}).
 An \textit{anti-isomorphism} between algebras $A$ and $A^\prime$ over $\mathbb{C}$
 is an anti-linear bijection $\alpha : A \to A^\prime$
 that also satisfies $\alpha (ab) = \alpha(b) \alpha(a)$ for all $a,b \in A$.
 For example, the next result immediately implies that the conjugation is an anti-isomorphism of the
 algebra $PG_{l,q}$ with itself for non-zero real $q$.
 
\begin{prop}
The algebra  $PG_{l,q}$ is a $*$-algebra, that is we have
$$
(fg)^* = g^*f^*
$$
for all $f,g \in  PG_{l,q}$, if and only if $q \in \mathbb{R} \setminus \{ 0 \}$.
\end{prop}
\textbf{Proof}: 
First take $q \ne 0$ and real.
We then consider the special case $f = \theta^i \overline{\theta}{}^j$ and  $g = \theta^k \overline{\theta}{}^m$,
where $i,j,k,m \in I_l$.
Then on one hand we have
\begin{gather*}
(f g)^* = (\theta^i \overline{\theta}{}^j  \theta^k \overline{\theta}{}^m)^* =
               (\theta^i q^{-jk} \theta^k \overline{\theta}{}^j  \overline{\theta}{}^m)^* =
               q^{-jk}  (\theta^{i+k} \overline{\theta}{}^{j+m} )^* 
               =  q^{-jk} \theta^{j+m} \overline{\theta}{}^{i+k} ,
\end{gather*}
while on the other we get
\begin{gather*}
g^* f^* = (\theta^k \overline{\theta}{}^m)^* (\theta^i \overline{\theta}{}^j)^* 
=  \theta^m \overline{\theta}{}^k  \theta^j \overline{\theta}{}^i
=  q^{-jk} \theta^{j+m}  \overline{\theta}{}^{i+k}.
\end{gather*}
So the result holds in this special case.
But since these are an arbitrary pair of elements in the basis $AW$, we get
$(f g)^* = g^* f^*$ for all $f, g \in PG_{l,q}$.

For $q \in \mathbb{C} \setminus \mathbb{R}$ equation (\ref{star-on-W-basis}) shows that
$PG_{l,q}$ is not a $*$-algebra.
$\quad \blacksquare$

\vskip 0.4cm \noindent
\textbf{Remarks}: 
In earlier preprint versions of this paper, I assumed that the conjugation made $PG_{l,q}$
into  a $*$-algebra.
So I only considered the case $q \in \mathbb{R} \setminus \{ 0 \}$.
This is an unnecessary restriction. 
Here we consider the more general case $q \in \mathbb{C} \setminus \{ 0 \}$. 
The only way that the conjugation enters into the subsequent theory is
through the definition of the sesquilinear form given in the next section.
And the properties that we will use of that sesquilinear form do not require that
the conjugation gives $PG_{l,q}$ a  $*$-algebra structure.

I thank R.~Fresneda \cite{fresneda} for clarifying for me that
the definition of the conjugation used in \cite{csq}  is the anti-linear extension of 
 $( \theta^i  \overline{\theta}{}^j )^*  =  \overline{\theta}{}^i \theta^j $.
(We note that this does not give a $*$-algebra.)
The fact that we will be using the above definition (\ref{define-star}) of the conjugation means that
we are considering structures that are strictly speaking distinct from those discussed in \cite{csq}.
Nonetheless, 
there will still be things in common with the approach in \cite{csq}.
We note that what is behind these various definitions of conjugation are different ways of dealing with an ordering problem.

The conjugation (\ref{define-star}) is nicely related to the integral.
The proof is left to the reader.
\begin{prop}
\begin{equation*}
 \int \! \! \! \int  d \theta \, f(\theta, \overline{\theta})^*  \, d \overline{\theta} =
 \left( \int \! \! \! \int  d \theta \, f(\theta, \overline{\theta})  \, d \overline{\theta} \right)^*
\end{equation*}
for all $f(\theta, \overline{\theta})  \in PG_{q,l}(\theta, \overline{\theta})$.
Here the conjugation on the right side is complex conjugation of a complex number.
\end{prop}

We would also like to note that there are other reasonable definitions for a conjugation.
One of these for any $q \in \mathbb{C} \setminus \{ 0 \}$ is  the anti-linear extension of the
following definition on elements of $AW$: 
\begin{equation}
(\theta^i \overline{\theta}{}^j)^{*} := (q^*/q)^{i j/2} \, \theta^j \overline{\theta}{}^i
\end{equation}
for a fixed choice of the square root of the phase factor $q^*/q$.
This extends our definition for $q \in \mathbb{R} \setminus \{ 0 \}$ provided that
we take $1^{1/2}$ to be $1$.
Note that this conjugation is an involution, that is $f^{**} =f$, and satisfies equations (\ref{more-relations}).
However for $q \notin \mathbb{R}$ it does \textit{not} satisfy $ (fg)^* = g^* f^*$ for all $f,g$.
It has the virtue of acting on the basis $W$ in a similar way to its action on $AW$.

Note that $ \theta$ and $\overline{\theta} $ are a pair of conjugate complex variables,
that is, $ \theta^* = \overline{\theta} $ and $\overline{\theta}{}^* = \theta $ 
and the intersection of the two subalgebras generated by  $ \theta$ and by $\overline{\theta} $,
respectively, is simply the smallest it could possibly be: $\mathbb{C} 1$.
This is in close analogy with the pair of conjugate complex variables $z$ and $\overline{z}$ (which are \textit{functions}
$\mathbb{C} \to \mathbb{C}$ and \textit{not} points in $\mathbb{C}$)
as studied in complex analysis, where $z$ generates the algebra of holomorphic functions on $\mathbb{C}$
while $\overline{z}$ generates the algebra of anti-holomorphic functions.
Also the intersection of these algebras consists of the constant functions.
Notice how the non-commutative geometry approach of viewing elements of algebras
(in this case functions) as the primary objects
of study clarifies a common confusion even in this commutative example where one might not
otherwise understand how the complex plane (whose complex dimension is one) can support two independent
complex variables, neither of which is more `fundamental' than the other.
This short discussion motivates the definition of a \textit{variable} as any element in a unital algebra that is not
a scalar multiple of the identity element.
And then a pair of \textit{complex variables} in any unital $*$-algebra is defined as any pair of conjugate variables which generate
subalgebras with intersection $\mathbb{C} 1$.

\section{Sesquilinear form}

Much of the material in this section comes from the paper \cite{csq}, though we define a more general sesquilinear form.

We want to introduce a sesquilinear form on $PG_{l,q}$ in order to turn it into something like
an $L^2$ space.
We start with any element (a `positive weight') in $PG_{l,q}$ of the form
\begin{equation}
\label{define-weight}
 w = w(\theta, \overline{\theta}) = \sum_{m \in I_l}  w_{l-1-m} \theta^m \overline{\theta}{}^m,
\end{equation}
where $w_n > 0$ for all $n \in I_l$.
The strange looking way of writing the sub-index on the right side of this equation will be justified later on.
In \cite{csq} the authors take $w_n = [n]_q!$ which is a $q$-deformed factorial of the integer $n$.
(See \cite{csq} for their definitions.) 
In any case, this couples the `weight' factors $w_n$ with the deformation parameter $q$, which itself is coupled
in \cite{csq} with the nilpotency power $l$.
We have preferred to keep all of these parameters decoupled from one and other.

Take $f = f(\theta, \overline{\theta})$ and $ g = g(\theta, \overline{\theta})$ in $PG_{l,q}$.
Informally, we would like to define
the \textit{sesquilinear form} or \textit{inner product} as in \cite{csq} by
\begin{equation}
\label{define_form}
\langle f , g \rangle_w := \int \!\!\! \int d \theta : f(\theta, \overline{\theta})^* g(\theta, \overline{\theta}) 
                                                                 w(\theta, \overline{\theta}) : d \overline{\theta},
\end{equation}
where $ : \, \, : $ is the \textit{anti-Wick} (or \textit{anti-normal}) \textit{ordering}, that is, put all $\theta$'s to the
left and all $\overline{\theta}$'s to the right.
However, the anti-Wick ordering is only well defined on the space $\mathbb{C} \{ \theta, \overline{\theta} \}$
and does not pass to its quotient space $PG_{l,q}(\theta, \overline{\theta})$.
By the formal expression (\ref{define_form}) we really mean
\begin{equation}
\label{rigor-define_form}
\langle f , g \rangle_w := \sum_{m \in I_l} w_{l-1-m} \int \!\!\! 
\int d \theta   \,\, \theta^m
              : f(\theta, \overline{\theta})^* : \, : g(\theta, \overline{\theta}) : 
\overline{\theta}{}^m \,\,  d \overline{\theta},
\end{equation}
where the \textit{anti-Wick product} $\, : \cdot : \, : \cdot : \,$
is defined as the $\mathbb{C}$-bilinear extension of this defining formula for pairs
of basis elements in $AW$:
\begin{equation}
: \theta^a \overline{\theta}{}^b : \, : \theta^c \overline{\theta}{}^d : \,\,\, \equiv \,\,\, \theta^{a+c} \overline{\theta}{}^{b+d} .
\end{equation}
So the anti-Wick product is a well defined mapping from $PG_{l,q} \times PG_{l,q}$ to $PG_{l,q}$.

Clearly the expression in (\ref{rigor-define_form}) is anti-linear in $f$ and linear in $g$.
Also, by the definition of the integral on the right side of (\ref{rigor-define_form}), we have that $\langle f , g \rangle_w $ is
a complex number.
Even though the multiplication in  $PG_{l,q}$  is not commutative we have the following result.
\begin{prop}
The sesquilinear form (\ref{rigor-define_form}) is complex symmetric, that is we have that
$ \langle f , g \rangle_w^* = \langle g , f \rangle_w$ for all $f,g \in PG_{l,q}$.
\end{prop}
\textbf{Proof}:
The point here is that the anti-Wick product is commutative.
A consequence is that 
$$
      \langle \theta^a \overline{\theta}{}^b ,  \theta^c \overline{\theta}{}^d \rangle_w =
       \langle \theta^c \overline{\theta}{}^d ,  \theta^a \overline{\theta}{}^b \rangle_w.
$$
Also this is a real number, as we will show momentarily below.
Then the result follows immediately by expanding $f$ and $g$ in the basis $AW$.
$\quad \blacksquare$

\vskip 0.4cm \noindent
We also define
\begin{equation}
\label{define_pnorm}
\vert \vert f \vert \vert^2_w := \langle f , f \rangle_w.
\end{equation}
Note that the previous proposition implies that $\langle f , f \rangle_w$ is real.
As we will subsequently see, this real number can be negative.
So, in general, we are not defining $\vert \vert f \vert \vert_w$.

Let $ \theta^a \overline{\theta}{}^b$ and $ \theta^c \overline{\theta}{}^d$ be arbitrary elements in the basis $AW$, for
$a, b, c, d \in I_l$.
We now compute their inner product:
\begin{eqnarray}
\label{compute-inner-product}
 \langle \theta^a \overline{\theta}{}^b ,  \theta^c \overline{\theta}{}^d \rangle_w &=&
 \int  \! \! \! \int d \theta : (\theta^a \overline{\theta}{}^b)^* \theta^c \overline{\theta}{}^d w(\theta, \overline{\theta}) : 
d  \overline{\theta} \nonumber
\\
 &=& \sum_{n \in I_l}  w_{l-1-n}
 \int  \! \! \! \int d \theta \,\, \theta^{n}  : \theta^b \overline{\theta}{}^a: \, : \theta^c \overline{\theta}{}^d :
  \overline{\theta}{}^{n} \,\,
 d  \overline{\theta} \nonumber
 \\
 &=& \sum_n w_{l-1-n}  \int  \! \! \!  \int d \theta \, \theta^{b+c+n}
  \overline{\theta}{}^{a+d+n} d  \overline{\theta} \nonumber
  \\
  &=& \sum_n w_{l-1-n} \, \delta_{b+c+n,l-1} \, \delta_{a+d+n,l-1}   \quad \mathrm{(Kronecker~deltas)} \nonumber
  \\
  &=& \Big\{ \begin{array}{cc} w_{b+c} = w_{a+d} & \mathrm{if~} b+c = a+d \le l -1 \\
                                         0 & \mathrm{otherwise} \nonumber
              \end{array}  \Big\}
\\     
\label{inner-product-calculation}         
    &=& \delta_{a+d, b+c} \, \chi_l (a+d) \, w_{a+d},                             
\end{eqnarray}
where $\chi_l$ is the characteristic function of $I_l$, that is for every integer $k$ we put $\chi_l(k) = 1$ if $k \in I_l$
and $\chi_l(k) = 0$ if $k \notin I_l$.
As noted above, (\ref{inner-product-calculation}) is a real number.
Strictly speaking, we should define $w_n$ here for $n \ge l$. But the exact
definition is unimportant since the $\chi_l$ factor is zero in that case.

Having established the closed formula (\ref{inner-product-calculation}) for the inner product of any pair of
elements in the basis $AW$, we will never again have need to calculate a Berezin integral.
Another way of saying this is that we could have taken (\ref{inner-product-calculation}) as the definition
of the inner product (extending anti-linearly in the first argument and linearly in the second)
and dispensed with the Berezin integral entirely.

Now in general there are pairs such that $(a,b) \ne (c,d)$ but $b+c = a+d \le l-1$, so that
there are distinct basis elements in $AW$ which are not orthogonal. 
(For example, take $a=1, b=l -1, c=0, d=l -2$. These all lie in $I_l$ since $l \ge 2$.)
That is to say, the basis $AW$ is not orthogonal.
Similar calculations show that the basis $W$ also is not orthogonal.

In particular, $\langle \theta^a \overline{\theta}{}^b,  \theta^c \overline{\theta}{}^d \rangle_w  = 0$ if $a-b \ne c -d$.
For example, taking $b=d=0$ and $a \ne c$, we see that
$\langle \theta^a,  \theta^c \rangle_w = 0$.
Similarly, taking $a=c=0$ and $b \ne d$, gives us  $\langle \overline{\theta}{}^b, \overline{\theta}{}^d \rangle_w = 0$.
Finally, taking $b=c=0$ and $a \ge 1$ and $d\ge 1$ we have that 
$\langle \theta^a,  \overline{\theta}{}^d \rangle_w =0$.
All of this shows that the following set is orthogonal:
$$
    \{ 1, \theta, \theta^2 , \dots , \theta^{l -1}, \overline{\theta}, \overline{\theta}{}^{2} , \dots, \overline{\theta}{}^{l -1} \}.
$$
Also, from  (\ref{compute-inner-product}) we immediately get
\begin{equation}
\label{norm-on-basis}
|| \theta^a \overline{\theta}{}^b  ||_w^2 = \Big\{ \begin{array}{cc} w_{a+b} & \mathrm{if~} a+b \le l -1 \\
                                         0 & \mathrm{otherwise}
              \end{array}    \Big\}
              = \chi_l(a+b) \, w_{a+b}.
\end{equation}
This already shows that there are non-zero elements $f \in PG_{l,q}$ such that $|| f ||_w^2 =0$.
One simply takes $f = \theta^a \overline{\theta}{}^b$ with $a,b \in I_l$ and $a+b \ge l$.
(For example, $a=b=l-1$ will do, since $l \ge 2$.)

Taking $b=0$ (with $a \in I_l$)  and $a=0$  (with $b \in I_l$) respectively in
equation (\ref{norm-on-basis}) we get the formulas
$$
    || \theta^a  ||_w^2 = w_{a} \quad \mathrm{and} \quad ||  \overline{\theta}{}^b  ||_w^2  = w_{b}.
$$
The `nice' subindices of the $w$'s in these identities are the reason we took the unusual
looking convention for the subindices in the definition (\ref{define-weight}) of $w(\theta, \overline{\theta})$.

There are also always elements $f \in  PG_{l,q}$ such that $|| f ||_w^2 < 0$.
To show this explicitly we note that we have these formulas for any $l \ge 2$:
\begin{gather*}
\langle 1, 1 \rangle_w = w_0, \quad \quad
\langle \theta^{l-1} \overline{\theta}{}^{l-1} , \theta^{l-1} \overline{\theta}{}^{l-1} \rangle_w = 0,
\\
\langle 1, \theta^{l-1} \overline{\theta}{}^{l-1} \rangle_w =
\langle \theta^{l-1}   \overline{\theta}{}^{l-1} , 1 \rangle_w = w_{l-1}.
\end{gather*}
Consequently, taking
$f= \alpha 1 + \beta \theta^{l-1} \overline{\theta}{}^{l-1} $ for any $\alpha, \beta \in \mathbb{C}$,  we have  that
\begin{gather*}
\langle f ,f \rangle_w =
\langle \alpha 1 + \beta \theta^{l-1} \overline{\theta}{}^{l-1} , \alpha 1 + \beta \theta^{l-1} \overline{\theta}{}^{l-1} 
\rangle_w =
| \alpha |^2 w_0 + (\alpha ^* \beta + \alpha \beta^*) w_{l-1}
\\
= | \alpha |^2 w_0 + ( | \alpha + \beta |^2 - | \alpha |^2   - | \beta |^2 )  w_{l-1}.
\end{gather*}
Since $w_{l-1} > 0$ it is not difficult to find $\alpha$ and $\beta$ making this last expression
strictly less than zero.
For example, $\alpha = 1$ and $\beta < - w_0/ (2 w_{l-1})$ will do.

We now define $\phi_n(\theta, \overline{\theta}) := w_n^{-1/2} \theta^n$
for every $n \in I_l$.
(Recall that $w_n > 0$. So we take the positive square root of $w_n$.)
We also write $\phi_n(\theta) := w_n^{-1/2} \theta^n$, since this element does not `depend' on
$\overline{\theta}$, that is, it lies in the subalgebra generated by $\theta$ alone.
These vectors clearly form an orthonormal basis of $\mathcal{B}_{H}$.
But given the above calculations we can immediately say more.
\begin{prop}
$\mathcal{B}_{H}$, $\mathcal{B}_{AH}$ and $\mathcal{S}$ are
Hilbert spaces with respect to $\langle \cdot , \cdot \rangle_w $., which is a positive definite inner
product when restricted to any of these subspaces of $PG_{l,q}$.
Moreover, with respect to this inner product we have the following statements:
\begin{enumerate}

\item
 $\{ \phi_n \, \vert \, n \in I_l  \}$ is an orthonormal basis of $\mathcal{B}_{H}$ and $\mathrm{dim}_{\mathbb{C}} \, \mathcal{B}_{H} = l$.
 
 \item
$\{ \phi_n^* \, \vert \, n \in I_l  \}$ is an orthonormal basis of $\mathcal{B}_{AH}$
and $\mathrm{dim}_{\mathbb{C}} \, \mathcal{B}_{AH} = l$.

\item
$\{ \phi_n \, \vert \, n \in I_l  \} \cup \{ \phi_n^* \, \vert \, n \in I_l \setminus \{ 0\} \} $ is an orthonormal basis of $\mathcal{S}$
and $\mathrm{dim}_{\mathbb{C}} \, \mathcal{S} = 2l - 1$.
\end{enumerate}
\end{prop}
Note that $\phi_n^*(\theta, \overline{\theta}) = ( \phi_n(\theta, \overline{\theta}) )^* =  w_n^{-1/2} \overline{\theta}{}^n$ using
$w_n > 0$.
We also write this element as $\phi_n^*( \overline{\theta})$.

\section{Reproducing kernel for \\ the Segal-Bargmann space}

It is not possible to find an algebra of complex valued
functions which is an isomorphic copy of the commutative algebra $\mathcal{B}_H$.
We can see this is so by simply noting that $\theta \in  \mathcal{B}_H$ satisfies $\theta \ne 0$ since $l \ge 2$
and is nilpotent, namely, $\theta^l =0$.
But no non-zero complex valued function is nilpotent.
Similarly, the commutative algebra $\mathcal{B}_{AH}$ is not isomorphic to an algebra of functions.

Nonetheless $\mathcal{B}_H$ and $\mathcal{B}_{AH}$ are reproducing kernel Hilbert spaces, 
properly understood.
The classical theory of  reproducing kernel Hilbert spaces whose elements are functions
goes back to the seminal work of Bergman in the 20th century. (See \cite{berg} for example.)
Here we start with $\mathcal{B}_H$, the Segal-Bargmann space, which we will now write
as $\mathcal{B}_H(\theta)$ to indicate that the paragrassmann variable in this space is $\theta$. 
First, let us note that when we write an arbitrary element in this  Segal-Bargmann space uniquely as
$$
        f(\theta) = \sum_{j \in I_l} \lambda_j \theta^j,
$$
where $\lambda_j \in \mathbb{C}$ for all $j \in I_l$,
this really can be interpreted as a function of $\theta$.
In fact, if we let $f(x) =  \sum_{j =0 }^N \beta_j x^j \in \mathbb{C}[x]$ be an arbitrary polynomial in $x$,
an indeterminant, then we can define a \textit{functional calculus}
(where the `functions' are polynomials)
 of any element $a \in \mathcal{B}_H(\theta)$ 
precisely by \textit{defining} $ f(a)$ to be $\sum_{j =0 }^N \beta_j a^j$.
With no loss of generality, we can take $N \ge l-1$.
If we now take $\beta_j = \lambda_j$ for all $j \in I_l$ and any value whatsoever for $\beta_j$ for $j \ge l$,
then $f(\theta)$ is simply the arbitrary element $\sum_j \lambda_j \theta^j$ considered above.
The mapping from $\mathbb{C}[x] $, the algebra of polynomials $f(x)$ in $x$, to $\mathcal{B}_H(\theta)$ given
by $f(x) \mapsto f(\theta)$ is clearly an algebra morphism that is surjective.
This is standard material, but it aids us in considering
$f(\theta) = \sum_{j \in I_l} \lambda_j \theta^j$ in $ \mathcal{B}_H(\theta)$ and its corresponding
element $f(\eta) = \sum_{j \in I_l} \lambda_j \eta^j$ in $ \mathcal{B}_H(\eta)$ for another paragrassmann  variable $\eta$. 
We provide $PG_{l,q}(\eta, \overline{\eta})$ and its subspace  $ \mathcal{B}_H(\eta)$ with essentially
the same inner product as above, simply replacing $\theta, \overline{\theta}$ with $\eta, \overline{\eta}$ everywhere.

Also, we wish to emphasize that this functional calculus is what replaces in this context the concept of
 `evaluation at a point' in the usual theory of reproducing kernel Hilbert spaces of functions.
 
We would like to establish for every $f(x) \in \mathbb{C}[x]$ the \textit{reproducing formula}
\begin{equation}
\label{reproducing-formula}
    f(\theta) = \langle K(\theta, \eta) , f(\eta) \rangle_w,
\end{equation}
where the inner product here is roughly speaking
taken with respect to the variable $\eta$ (that is, basically in $\mathcal{B}_H(\eta)$; more on this later)
and where 
$$
K(\theta, \eta) \in  \mathcal{B}_{AH}(\overline{\theta}) \otimes \mathcal{B}_H(\eta).
$$
This last condition says that the \textit{reproducing kernel} $K(\theta, \eta)$ 
is holomorphic in $\eta$ and anti-holomorphic in $\theta$. 
This condition is in analogy with the theory of reproducing kernel functions in holomorphic
function spaces. (See \cite{hall}, for example.)
The reader should note that we are using the notation $K(\theta, \eta)$, which is analogous to the notation in the
classical theory of reproducing kernel Hilbert spaces. 
If we had been consistent with our own conventions, we would have denoted this as $K(\overline{\theta}, \eta)$.

Now the unknown we have to solve for is the kernel $K(\theta, \eta)$.
We write  
\begin{equation}
\label{general-K}
    K(\theta, \eta) = \sum_{i,j}   a_{ij} \overline{\theta}{}^i \otimes \eta^j,
\end{equation}
an arbitrary element in
$  \mathcal{B}_{AH}(\overline{\theta}) \otimes \mathcal{B}_H(\eta)$, and see what are the conditions that
the reproducing formula (\ref{reproducing-formula}) imposes on the coefficients
$a_{ij} \in \mathbb{C}$.
We take an arbitrary element $f(\theta) = \sum_k \lambda_k \theta^k \in \mathcal{B}_H (\theta)$.
So we have the corresponding element $f(\eta) = \sum_k \lambda_k \eta^k \in \mathcal{B}_H (\eta)$.
We then calculate out the right side of equation (\ref{reproducing-formula}) and get
\begin{gather*}
   \langle K(\theta, \eta) , f(\eta) \rangle_w = 
   \langle \sum_{i,j}   a_{ij} \, \overline{\theta}{}^i \otimes \eta^j , f(\eta) \rangle_w
   = \sum_{i,j}   a_{ij}^* \,  \langle \eta^j , f(\eta) \rangle_w \, \theta^i 
   \\
   = \sum_{i,j} a_{ij}^* \,   \langle \eta^j , 
        \sum_k \lambda_k \eta^k \rangle_w \, \theta^i 
         = \sum_{i,j} a_{ij}^* \,  \sum_k \lambda_k \langle \eta^j , 
        \eta^k \rangle_w \, \theta^i 
   \\
         = \sum_{i,j} a_{ij}^* \,   \sum_k \lambda_k \delta_{j,k} w_j \, \theta^i 
         = \sum_{i,j} a_{ij}^* \,   \lambda_j w_j \, \theta^i 
          = \sum_i \left( \sum_j w_j a_{ij}^* \, \lambda_j \right) \theta^i  .
\end{gather*}
(Note that the second equality is nothing other than the promised, rigorous definition of the inner product
in (\ref{reproducing-formula}).)
Now we want this to be equal to $f(\theta) = \sum_i \lambda_i \theta^i$
for \textit{all} $f(\theta)$ in $\mathcal{B}_H (\theta)$, that is, for all vectors $\{ \lambda_i | i \in I_l \}$ in $\mathbb{C}^l$.
So the matrix $(w_j a_{ij}^*)$ has to act as the identity on $\mathbb{C}^l$ and thus has to be the
identity matrix $(\delta_{ij})$, where $\delta_{ij}$ is the Kronecker delta.
The upshot is that $a_{ij} = \delta_{ij} / w_j$ does the job, and nothing else does.
So, substituting in equation (\ref{general-K}) we see that
\begin{equation}
\label{reproducing-kernel}
         K(\theta, \eta) = \sum_{j} \dfrac{1}{w_j} \overline{\theta}{}^j \otimes \eta^j
\end{equation}
is the unique reproducing kernel `function.'

And, as one might expect, an abstract argument also shows that reproducing
kernels are unique.
For suppose that $K_1(\theta, \eta) \in \mathcal{B}_{AH}(\overline{\theta}) \otimes  \mathcal{B}_{H}(\eta)$ is also a reproducing kernel.
Then the standard argument makes sense in this context, namely,
\begin{gather}
       K_1(\rho, \eta) = \langle K(\eta, \theta) , K_1(\rho, \theta) \rangle_w = 
       \langle K_1(\rho, \theta) , K(\eta, \theta)  \rangle_w^* \nonumber
       \\
       \label{kernel-unique}
        = K(\eta, \rho) ^* =  K(\rho, \eta),
\end{gather}
where $\rho$ is another paragrassmann variable.
The astute reader will have realized that these innocent looking formulas require a bit
of justification, including a rigorous definition of the inner product in this context.
We leave most of these details to the reader.
But, for example, in the last equality we are using the standard natural isomorphisms
$$
\left( \mathcal{B}_{AH}(\overline{\eta}) \otimes \mathcal{B}_H(\rho) \right)^*  
\cong  \mathcal{B}_{H}(\eta) \otimes \mathcal{B}_{AH}(\overline{\rho})
\cong \mathcal{B}_{AH}(\overline{\rho}) \otimes  \mathcal{B}_{H}(\eta).
$$
The relation $K(\eta, \rho) ^* =  K(\rho, \eta)$ is also well known in the classical theory.
For example,  Eq. (1.9a) in \cite{BA} is an analogous result.
Also, by putting $K_1$ equal to $K$ in the first equality of (\ref{kernel-unique}) we get another result that
is analogous to a result in the classical case.
This is
\begin{equation}
\label{inner-product-of-two-Ks}
  K(\rho, \eta) = \langle K(\eta, \theta) , K(\rho, \theta) \rangle_w, 
\end{equation}
which is usually read as saying that two `evaluations' of the reproducing kernel
(with the holomorphic variable, here $\theta$, being the same in the two) have an inner product with respect to
that holomorphic variable that is itself an `evaluation' of the reproducing kernel.
See \cite{hall}, Theorem 2.3, part 4, for the corresponding identity in the context of
holomorphic function spaces.
Next we put $\eta = \rho$ to get
\begin{equation}
\label{K-rho-rho}
  K(\rho, \rho) = \langle K(\rho, \theta) , K(\rho, \theta) \rangle_w = ||  K(\rho, \theta) ||_w^2 = ||  K(\rho, \cdot) ||_w^2.
\end{equation}
And again this last formula is analogous to a result in the classical theory.
Later on, we will discuss the positivity of $K(\rho, \rho)$.

We note that this theory is consistent with many expectations coming from the usual theory of reproducing kernels.
As we have seen, $K(\theta, \eta)$ is the only element in the appropriate space with the reproducing property.
Also, we clearly have the following well known relation with the elements of the standard orthonormal basis, namely that
\begin{equation}
\label{onb-expansion-1}
     K(\theta, \eta) = \sum_{j} \phi_j (\theta)^* \otimes \phi_j (\eta).
\end{equation}
This follows from (\ref{reproducing-kernel}) and the definition of $\phi_j$.
Suppose that $\psi_j(\eta)$ for $j \in I_l$ is another orthonormal basis of $\mathcal{B}_{H}(\eta)$.
It then follows from 
$$
    \langle f^*, g^* \rangle_w =  \langle f, g \rangle_w^*
$$
for all $f,g \in PG_{l,q}$ 
(which we leave to the reader as another exercise)
that $\psi_j(\theta)^*$
is an orthonormal basis of $\mathcal{B}_{AH}(\theta)$.
Then by a standard argument in linear algebra we obtain from (\ref{onb-expansion-1}) that
$$ 
     K(\theta, \eta) = \sum_{j} \psi_j (\theta)^* \otimes \psi_j (\eta).
$$
This formula is the analogue of an identity in the classical theory.
(For example see \cite{BA}, Eq. (1.9b) or \cite{hall}, Theorem 2.4.)

One result from the classical theory of reproducing kernel Hilbert spaces of functions seems to have no analogue in
this context.
That result is the point-wise estimate that one gets by applying the Cauchy-Schwarz inequality
to the reproducing formula. 
(For example, see \cite{hall}, Theorem 2.3, part 5 for this result in the holomorphic function setting.)
What happens in the classical case is that one takes the absolute value of an expression
such as $f(z)$ which is a complex number (being the value of a function $f$ at the point $z$) and
then applies Cauchy-Schwarz to the inner product on the right side of  the reproducing formula,
which gives a bound by the product of two Hilbert space norms: one of the reproducing kernel with
respect to its holomorphic variable and the other of $f$.
So there are two semi-norms of $f$ in this formulation.
The first is $|f(z)|$ and the second (actually a norm) is $|| f ||$ in the appropriate Hilbert space norm.
But in the context of this paper the expression $f(\theta)$ on the left side of (\ref{reproducing-formula})
is an element in the Hilbert space $\mathcal{B}_H(\theta)$ with dimension greater than one; in particular
it is not a complex number.
Moreover, the expression $f(\eta)$ on the right side of (\ref{reproducing-formula}) is an element
in the isomorphic Hilbert space $\mathcal{B}_H(\eta)$ and as such is the isomorphic copy of $f(\theta)$ 
in $\mathcal{B}_H(\theta)$.
So $f(\theta)$ and $f(\eta)$ are essentially the same object, and it seems that we have available only one
semi-norm to measure each of them, namely the norm in the corresponding Hilbert space.
Of course, by definition of the norms 
$$
      || f(\theta) ||_{\mathcal{B}_H(\theta)} =  || f(\eta) ||_{\mathcal{B}_H(\eta)}
$$
holds.
One could weaken this to an estimate

$$
      || f(\theta) ||_{\mathcal{B}_H(\theta)} \le  || f(\eta) ||_{\mathcal{B}_H(\eta)}
$$
and say that this is the point-wise estimate in this context.
Naturally, saying such a thing is rather ridiculous, albeit true.
So, while the discussion of this paragraph does not preclude the possibility of another
semi-norm that gives a non-trivial `point-wise' estimate, it does indicate what is behind this issue.
However, I have not been able to find such a suitable semi-norm.

The reproducing formula in the usual theory of reproducing kernel Hilbert spaces is  interpreted as 
saying that the Dirac delta function is realized as integration against  a smooth kernel function.
Since the inner product in the reproducing formula (\ref{reproducing-formula}) is a Berezin integral, we can
say that the Dirac delta function in this context is realized as Berezin integration against the `smooth
function' $K(\theta, \eta)$.
But what is the Dirac delta function in this context?
We recall that $f(\theta)$ in this paper is merely convenient notation for an element in
a Hilbert space. We are not evaluating $f$ at a point $\theta$ in its domain.
It seems that the simplest interpretation for the Dirac delta $\delta_{\eta \to \theta}$ in the present context is that
it acts on $f(\eta)$ to produce $f(\theta)$, namely that it is a \textit{substitution} operator.
We use a slightly different notation for this Dirac delta in part because it is different from the usual
Dirac delta  and also to distinguish it from the Kronecker delta function that we have been using.
An appropriate definition and notation would be $\delta_{\eta \to \theta} [ f(\eta)] := f(\theta)$ so that
$ \delta_{\eta \to \theta} : \mathcal{B}_H (\eta) \to \mathcal{B}_H (\theta)$
is an isomorphism of Hilbert spaces and of algebras.
In particular, with this way of defining the Dirac delta we do not get a functional acting on a space
of test functions.
However, the left side of equation (\ref{reproducing-formula})  is $\delta_{\eta \to \theta} [ f(\eta)] $.
Notice that in this approach 
$$  \delta_{\eta \to \theta} \in 
\mathrm{Hom}_{\mathrm{Vect}_\mathbb{C}} ( \mathcal{B}_H (\eta) , \mathcal{B}_H (\theta) ) \cong
\mathcal{B}_H (\theta) \otimes \mathcal{B}_H (\eta)^\prime \cong \mathcal{B}_H (\theta) \otimes \mathcal{B}_{AH}  (\overline{\eta}) 
$$
using standard notation from category theory and viewing the space $\mathcal{B}_{AH}  (\overline{\eta})$
as the dual space $ \mathcal{B}_H (\eta)^\prime$.
This does agree with our previous analysis where we had that
$K(\theta, \eta) \in \mathcal{B}_{AH}(\overline{\theta}) \otimes \mathcal{B}_H(\eta)$, because 
the inner product in equation (\ref{reproducing-formula}) is anti-linear in its first argument.
So that previous analysis simply identifies which element in
$\mathcal{B}_{H} (\theta) \otimes \mathcal{B}_{AH}  (\overline{\eta}) $
is the Dirac delta, namely
$$
 \delta_{\eta \to \theta} =  \sum_{j} \dfrac{1}{w_j} \theta^j \otimes \overline{\eta}{}^j = \sum_{j} \phi_j (\theta)
 \otimes \phi_j^* (\overline{\eta}),     
$$
which is analogous to a standard formula for the Dirac delta.

We now try to see to what extent this generalized notion of a reproducing kernel has the positivity
properties of a usual reproducing kernel. First, we `evaluate' the diagonal `elements' $K(\theta, \theta)$
in $\mathcal{B}_{AH}(\overline{\theta}) \otimes \mathcal{B}_{H}(\theta)
\cong
\mathcal{B}_{H}(\theta)^\prime \otimes \mathcal{B}_{H}(\theta)$
getting
\begin{gather*}
  K(\theta, \theta) = \sum_j \dfrac{1}{w_j} \overline{\theta}{}^j \otimes \theta^j
=  \sum_j \dfrac{1}{w_j} (\theta^j)^* \otimes \theta^j.
\end{gather*}
This is  positive element by using the usual definition of a positive
element in a $*$-algebra, \textit{provided} we adequately define the $*$-operation in 
$ \mathcal{B}_{AH}(\overline{\theta}) \otimes \mathcal{B}_{H}(\theta)
\cong \mathcal{B}_{H}(\theta)^\prime \otimes \mathcal{B}_{H}(\theta) $. 
Given that this exercise has been done and
without going into further details we merely comment that we can identify
$\mathcal{B}_{H}(\theta)^\prime \otimes \mathcal{B}_{H}(\theta)$ with $\mathcal{L} (\mathcal{B}_{H}(\theta) )$,
the vector space (and $*$-algebra) of all of the linear operators from the Hilbert space $\mathcal{B}_{H}(\theta)$ to itself.
Under this identification the positive elements of $\mathcal{B}_{H}(\theta)^\prime \otimes \mathcal{B}_{H}(\theta)$
correspond exactly to the positive operators in $\mathcal{L} (\mathcal{B}_{H}(\theta) )$, and we have that
$$
  K(\theta, \theta) = 
  \sum_j \dfrac{1}{w_j}   | \theta^j \rangle \langle \theta^j | = \sum_j  | \phi_j (\theta) \rangle \langle \phi_j (\theta) |
  \in \mathcal{L} (\mathcal{B}_{H}(\theta) ),
$$
using the Dirac bra and ket notation.
But this is clearly a positive linear operator since 
$$
       \sum_j  | \phi_j (\theta) \rangle \langle \phi_j (\theta) | = I_{ \mathcal{B}_{H}(\theta) } \equiv I \ge 0 ,
$$
the identity operator on $\mathcal{B}_{H}(\theta)$.
The upshot is that $ K(\theta, \theta) = I$.
But $||  K( \theta, \cdot ) ||_w^2 =  K( \theta, \theta)$ as we showed in (\ref{K-rho-rho}).
So, $||  K( \theta, \cdot ) ||_w^2 = I$ as well.

Next, we take a finite number of pairs of paragrassmann variables $\theta_n, \overline{\theta}_n$ and a finite sequence of
complex numbers $\lambda_n$,
where $n = 1, \dots , N$.
Then we investigate the positivity of the usual expression, that is, we consider
\begin{gather*}
\sum_{n,m=1}^N \lambda_n^* \lambda_m K(\theta_n , \theta_m) = 
\sum_{n,m=1}^N \lambda_n^* \lambda_m \sum_{j \in I_l} \dfrac{1}{w_j} \overline{\theta}{}^j_n \otimes \theta_m^j
\\
=  \sum_{j \in I_l} \dfrac{1}{w_j} \sum_{n,m=1}^N \lambda_n^* \lambda_m (\theta_n^j)^*
\otimes \theta_m^j
=
\sum_{j \in I_l} \dfrac{1}{w_j} \left(   \sum_{n=1}^N  \lambda_n \theta_n^j \right)^*  
                                                        \otimes \left( \sum_{m=1}^N  \lambda_m \theta_m^j
                                              \right),
\end{gather*}
which is a positive element in the appropriate $*$-algebra and therefore also
corresponds to a positive linear map.
A detail here is that one has to define a $*$-algebra where the sums $\sum_{n}  \lambda_n \theta_n^j$
makes sense for all $j \in I_l$.
But this is a straightforward exercise left to the reader.

It now is natural to ask whether this procedure can be reversed, as we know is the case with the usual theory
of reproducing kernel functions.
That is to say, can we start with a mathematical object, call it $K$, that has the properties
(in particular, the positivity) of a reproducing kernel in this context and produce from it
a reproducing kernel Hilbert space that has that given object $K$ as its reproducing kernel?
This seems not to be possible, at least not using an argument based on the identity
(\ref{inner-product-of-two-Ks}) as is done in the classical case.
It turns out that (\ref{inner-product-of-two-Ks}) is only analogous to the identity in the classical case,
since it says something decidedly different given that
the left side of it is not a complex number.
In the classical theory the operation of evaluation at a point gives a complex number.
But in this context the evaluation of a function at a variable in an algebra gives another
element in that same algebra.
In the argument in the classical case, one uses the analogue of (\ref{inner-product-of-two-Ks})
to define an inner product (on a set of functions) having available only the candidate mathematical object $K$
(in that case a function of two variables).
But one can not use (\ref{inner-product-of-two-Ks}) directly to define in this context a complex valued inner product.
Perhaps an inverse procedure can be found, but it will have to differ somewhat from the procedure
in the classical case.

We now come back to the question of finding an analogy to a point-wise bound for $f (\theta) \in \mathcal{B}_H(\theta)$.
Suppose that $f (\theta) \ne 0$ and put $u_0 := || f(\theta) ||_w^{-1} f(\theta)$,
a unit vector in $ \mathcal{B}_H(\theta)$.
Extend this to an orthonormal basis $u_j$ of $ \mathcal{B}_H(\theta)$  for $j \in I_l$.
So we get the operator inequality
$ | u_0 \rangle \langle u_0 | \le \sum_{j} | u_j \rangle \langle u_j | = I_{\mathcal{B}_H(\theta)}$.
Then we obtain
\begin{equation}
\label{pw}
| f(\theta) \rangle \langle f(\theta) | = || f(\theta) ||_w^2  \, | u_0 \rangle \langle u_0 |
 \le  || f(\theta) ||_w^2 \, I_{\mathcal{B}_H(\theta)} =
|| f(\theta) ||_w^2 \, ||  K( \theta, \cdot ) ||_w^2
\end{equation}
which is an operator inequality involving positive operators.
For $f(\theta) = 0$
this inequality is trivially true (and is actually an equality).
The point here is that (\ref{pw}) has some resemblance to the point-wise 
estimate in the classical case. (See \cite{hall}.)
The left side of (\ref{pw}) can be considered a type of  `outer product'  of $f(\theta)$ with itself.

An entirely analogous argument shows that  $\mathcal{B}_{AH}$ is a reproducing kernel Hilbert space
with reproducing kernel $K_{AH}$ given by
$$
     K_{AH}(\theta, \eta) = K(\theta, \eta)^* =  \sum_{j} \phi_j (\theta) \otimes \phi_j (\eta)^* =
     \sum_{j} \dfrac{1}{w_j} \theta^j \otimes \overline{\eta}{}^j.
$$
The theory for this reproducing kernel is really the same as the material presented in this section.

Finally, we note that the results of this section depend on the weight $w(\theta, \overline{\theta})$ and
are independent of the value of the parameter $q \in \mathbb{C} \setminus \{ 0 \}$.

\section{Coherent States and the \\ Segal-Bargmann Transform}

We now introduce the coherent state quantization of Gazeau.
(See \cite{jpg} for example.)
We let $\mathcal{H}$ be any complex Hilbert space of dimension $l$ and choose any
orthonormal basis of  $\mathcal{H}$, which we will denote as  $e_n$ for $n \in I_l$.
While $\theta$ is a complex variable, it does not `run' over a domain of values, say in some phase space.
So the coherent state $| \theta \rangle $ we are about to define is one object, not a parameterized family of objects.
Actually, we define two \textit{coherent states} corresponding to the variable $\theta$:
$$
| \theta \rangle := \sum_{n \in I_l} \phi_n (\theta) \otimes e_n \in \mathcal{B}_{H} (\theta)
\otimes \mathcal{H},
$$
$$
\langle \theta | := \sum_{n \in I_l} \phi_n^* (\overline{\theta}) \otimes e_n^\prime \in \mathcal{B}_{AH} (\theta)
\otimes \mathcal{H}^\prime.
$$
Here $ \mathcal{H}^\prime$ denotes the dual space of all linear functionals on $ \mathcal{H}$,
and  $e_n^\prime$ is its orthonormal basis that is dual to $e_n$, where $n \in I_l$.
We refrain from using the super-script $*$ for the dual objects in order to avoid
confusion in general with the conjugation.
We are following Gazeau's conventions here.
(See \cite{jpg}.)
As noted in \cite{csq}
these objects give a resolution of the identity using the Berezin integration theory,
and so this justifies calling them coherent states.
Now we have
\begin{gather*}
\langle \, \theta \, | \, \eta \, \rangle = \sum_{j,k} \phi_j^* (\overline{\theta}) \otimes \phi_k (\eta) \,
\langle e_j^\prime , e_k \rangle = 
 \sum_{j,k} \phi_j^* (\overline{\theta}) \otimes \phi_k (\eta) \, \delta_{j,k}
 \\
=  \sum_{j} \phi_j^* (\overline{\theta}) \otimes \phi_j (\eta) = K(\theta,\eta).
\end{gather*}
Here the inner product  (or  pairing) of the coherent states $\langle \, \theta \, | $ and
$| \, \eta \, \rangle$ is simply \textit{defined} by the first equality, which is a quite natural definition.
So the inner product of two coherent states gives us the reproducing kernel.
This is a result that already appears in the classical theory. (See \cite{BA}, Eq. (1.9a) for a formula of this type.)

Since we have coherent states, we can define the corresponding \textit{Segal-Bargmann}
(or \textit{coherent state}) transform in the usual way.
Essentially the same transform was discussed in \cite{csq}, where it is denoted by $\mathcal{W}$.
This will be a unitary isomorphism $C: \mathcal{H} \to \mathcal{B}_{H}(\theta)$.
(Recall that we are considering an abstract Hilbert space $\mathcal{H}$ of dimension~$l$ with orthonormal basis $e_n$.)
We define $C$ for all $\psi \in \mathcal{H}$ by
\begin{equation}
\label{define-cst}
    C \psi (\theta) := \langle \, \langle \theta |, \psi \rangle.
\end{equation}
Note that $ \langle \theta | \in \mathcal{B}_{AH}(\theta) \otimes \mathcal{H}^\prime$.
So the outer bracket $\langle \cdot , \cdot \rangle$ here refers to the pairing of the dual space
$\mathcal{H}^\prime$ with $\mathcal{H}$.
As noted earlier $ \langle \theta |$ is one object, not a parameterized family of objects.
So $ C \psi (\theta)$ is one element in the Hilbert space $\mathcal{B}_{H}(\theta)$. And as usual
the $\theta$ in the notation is a convenience for reminding us what the variable is, and it is not a
point at which we are evaluating $C \psi$.
Actually, in this context where it is understood that the variable under consideration is $\theta$,
$ C \psi (\theta)$ and $ C \psi$ are two notations for one and the same object.

Substituting the definition for $\langle \theta |$ gives
\begin{gather*}
          C\psi (\theta) = \langle \, \langle \theta |, \psi \rangle
                               = \left \langle \sum_{n \in I_l} \phi_n^* (\overline{\theta}) \otimes e_n^\prime , \psi \right\rangle
                               \\
                               =  \sum_{n \in I_l}  \langle \phi_n^* (\overline{\theta}) \otimes e_n^\prime , \psi \rangle
                               =  \sum_{n \in I_l} \langle e_n^\prime , \psi \rangle \, \phi_n (\theta) .
\end{gather*}
So by taking $\psi$ to be $e_j$ we see that the Segal-Bargmann transform
$C: \mathcal{H} \to \mathcal{B}$ is the (unique) linear transformation mapping the orthonormal basis 
$e_j$ to the orthonormal basis $\phi_j(\theta)$.
This shows that $C$ is indeed a unitary isomorphism.
Also the above shows that this is an `integral kernel' operator with kernel given by the coherent state
$ \langle \theta | = \sum_j   \phi_j^*(\overline{\theta})  \otimes e_j^\prime$.
This is also a well-known relation in the Segal-Bargmann  theory.
Equation (2.10b)  in \cite{BA} is this sort of formula.

The Segal-Bargmann transform of the coherent state $| \, \eta \, \rangle$
is given by 
$$
C | \, \eta \, \rangle (\theta) = \langle \, \langle \theta |, | \, \eta \, \rangle \rangle = 
\langle \theta \, | \, \eta \, \rangle = K(\theta, \eta).
$$
So, this says that the Segal-Bargmann transform of a coherent state is the reproducing kernel.
This is also a type of relation known in the classical context of \cite{BA}, where it appears as Eq. (2.8).

There are also two coherent states corresponding to the other variable $\overline{\theta}$, which
also give a resolution of the identity.
And $\langle \, \overline{\theta} \, | \, \overline{\eta} \, \rangle$ gives the reproducing kernel for the
anti-holomorphic Segal-Bargmann space, and so forth.

\section{Reproducing kernel for the \\ Paragrassmann space}

Consider $PG_{l,q}(\theta, \overline{\theta})$ with $l=2$, a `fermionic' case.
This is a non-commutative algebra of dimension $4$.
The basis $AW$ is $\{ 1, \theta,  \overline{\theta} , \theta \overline{\theta} \}$.
The weight `function' in this case is
$$
     w(\theta, \overline{\theta}) = w_1 + w_0 \theta \overline{\theta}.
$$
(We are using our convention for the sub-indices. See equation (\ref{define-weight}).)
So if this space has a reproducing kernel $K_{PG}  (\theta, \overline{\theta}, \eta, \overline{\eta} )$,
it must satisfy
$$
f  (\theta, \overline{\theta} ) = \langle K_{PG}  (\theta, \overline{\theta}, \eta, \overline{\eta} ) , f  (\eta, \overline{\eta} ) \rangle_w
$$
for all $f = f  (\theta, \overline{\theta} ) \in PG_{2,q}$.
(Notice that here, and throughout this section, we                                                                                                                                                                                                                                                                                                                                                                                                                                                                                                                                 use a functional calculus of a pair of \textit{non-commuting} variables in place of evaluation at a point.
Actually, we can define a functional calculus for all $f \in \mathbb{C} \{ \theta, \overline{\theta} \}$.)
This last equation in turn is equivalent to satisfying
$$
\theta^i \overline{\theta}{}^j = 
\langle K_{PG}  (\theta, \overline{\theta}, \eta, \overline{\eta} ) , \eta^i \overline{\eta}^j \rangle_w
$$
for all $i,j \in I_2 = \{ 0, 1 \}$.
Substituting the general element 
$$
K_{PG}  (\theta, \overline{\theta}, \eta, \overline{\eta} ) = \sum_{abcd} k_{abcd} \theta^a \overline{\theta}{}^b \otimes
\eta^c \overline{\eta}{}^d
$$ 
(which lies in a space of dimension $2^4$) into the previous equation gives
$$
\theta^i \overline{\theta}{}^j = 
\sum_{abcd}   k_{abcd}^*  \theta^b \overline{\theta}{}^a  \langle \eta^c \overline{\eta}{}^d , \eta^i \overline{\eta}^j \rangle_w
=
\sum_{ab} \left(   \sum_{cd} k_{abcd}^* G_{cdij} \right) \theta^b \overline{\theta}{}^a ,
$$
where $ G_{cdij} = \langle \eta^c \overline{\eta}{}^d , \eta^i \overline{\eta}^j \rangle_w$.
So the unknown coefficients $ k_{abcd} \in \mathbb{C}$ must satisfy
$$
       \sum_{cd} k_{abcd}^* G_{cdij} = \delta_{(b,a),(i,j)},
$$
where this Kronecker delta is $1$ if the ordered pair $(b,a)$ is equal to the ordered pair $(i,j)$ and
otherwise is $0$.
So everything comes down to showing the invertibility of the matrix $ G = ( G_{cdij} )$,
whose rows (resp., columns) are labelled by $\eta^c \overline{\eta}{}^d$ (resp., $\eta^i \overline{\eta}^j \rangle_w$)
where the ordered pairs $(c,d), (i,j) \in I_2 \times I_2$.
So $G$ is a $4 \times 4$ matrix.
To calculate it we note that
the pertinent identities are as follows:
\begin{gather*}
\langle 1 ,1\rangle_w = w_0,
\\
\langle 1 , \eta \overline{\eta} \rangle_w = \langle  \eta \overline{\eta} , 1\rangle_w = w_1,
\\
\langle \eta, \eta\rangle_w = \langle \overline{\eta} , \overline{\eta} \rangle_w = w_1,
\\
\langle \eta \overline{\eta} , \eta  \overline{\eta} \rangle_w = 0.
\end{gather*}
All other inner products of pairs of elements in $AW$ are zero.
The matrix $G$ we are dealing with here in the case $l=2$ is
\begin{equation*}
G = \left(
        \begin{array}{cccc}
        w_0 &  0 &  0 &  w_1 \\
        0   & w_1 & 0 & 0   \\
        0  &  0  &   w_1 &  0 \\
        w_1 & 0 & 0 & 0
        \end{array}
        \right)
\end{equation*}
with respect to the ordered basis $\{ 1 , \eta, \overline{\eta}, \eta \overline{\eta} \}$.
Then $\det G = - (w_1)^4 \ne 0$ and so
\begin{equation*}
G^{-1} = \left(
        \begin{array}{cccc}
        0 &  0 &  0 &  1/w_1 \\
        0   & 1/w_1 & 0 & 0   \\
        0  &  0  &   1/w_1 &  0 \\
        1/w_1 & 0 & 0 & -w_0/w_1^2
        \end{array}
        \right)
\end{equation*}
by using standard linear algebra.
It follows that the (unique!) reproducing kernel for $PG_{2,q}$ is given by
\begin{equation}
\label{KPG-theta-eta}
K_{PG} (\theta, \overline{\theta}, \eta, \overline{\eta} ) = \dfrac{1}{w_1} \theta \overline{\theta} \otimes 1
+  \dfrac{1}{w_1} \overline{\theta} \otimes \eta  + \dfrac{1}{w_1}  \theta \otimes \overline{\eta} + 
\dfrac{1}{w_1} 1 \otimes \eta  \overline{\eta} 
- \dfrac{w_0}{w_1^2}  \theta \overline{\theta} \otimes  \eta \overline{\eta} .
\end{equation}
(This is also works when $w_0 \le 0$ or $w_1 < 0$.)
Even though the reproducing kernel in this example lies in a space of dimension $16$,
only $5$ terms in the standard basis have non-zero coefficients.

Actually, the method in the previous paragraph is the systematic way to arrive at a formula
for the reproducing kernel for $PG_{l,q}$ in general.
Everything comes down to showing the invertibility of the matrix $G$, where
\begin{equation}
\label{matrix-G}
 G_{cdij} = \langle \eta^c \overline{\eta}{}^d , \eta^i \overline{\eta}^j \rangle_w
 \end{equation}
for $(c,d),(i, j) \in I_l \times I_l$ and then finding the inverse matrix.
Again, we label the rows and columns of $G$ by the elements in $AW$.
As is well known, invertibility is a generic property of $G$ (that is, true for an open, dense set of matrices $G$).
So there are many, many examples of sesquilinear forms (including positive definite inner products) defined
on the non-commutative algebra $PG_{l,q}$, making it into a reproducing kernel space.
Then it becomes clear that it is straightforward to give any finite dimensional
algebra $\mathcal{A}$, commutative or not  (but such that every element in $\mathcal{A}$ is
in the image of some functional calculus), an inner product  so that $\mathcal{A}$ has a reproducing kernel.
The infinite dimensional case will require more care due to the usual technical details.

But is the matrix $G$ associated to 
the sesquilinear form (\ref{rigor-define_form}) invertible?

\begin{theorem}
\label{det-of-G}
Taking $G$ to be the matrix (\ref{matrix-G}) associated to the  sesquilinear form defined by (\ref{rigor-define_form}),
we have that
$\det G = \pm (w_{l-1})^{l^2} \ne 0$ for every $l \ge 2$.
\end{theorem}
\textbf{Proof}: We will argue by induction on $l$ for $l \ge 4$.

We have shown above that $\det G = - (w_1)^4 = - (w_1)^{2^2} \ne 0$ when $l=2$.
So we must establish this result for $l=3$ as well. We claim in that case
that $\det G = (w_2)^9 \ne 0$.
First we calculate the matrix entries of $G$, which is a $9 \times 9$ matrix, and get
\begin{equation*}
G =  \left( \begin{array}{ccccccccc}
                        w_0 & 0 & 0 & w_1 & 0 & 0 & 0 & 0 & w_2 \\
                        0 & w_1 & 0 & 0 & 0 & 0 & 0 & w_2 & 0 \\
                        0 & 0 & w_1 & 0 & 0 & 0 & w_2 & 0 & 0 \\
                        w_1 & 0 & 0 & w_2 & 0 & 0 & 0 & 0 & 0 \\
                        0 & 0 & 0 & 0 & 0& w_2 & 0 & 0 & 0 \\
                        0 & 0 & 0 & 0 & w_2 & 0& 0 & 0 & 0 \\
                        0 & 0 & w_2  & 0 & 0 & 0 & 0 & 0 & 0 \\
                        0 & w_2  & 0 & 0 & 0 & 0 & 0 & 0  & 0\\
                        w_2  & 0 & 0 & 0 & 0 & 0 & 0  & 0 & 0 \\
                 \end{array}
         \right)
\end{equation*}
with respect to the ordered basis
$\{  1, \theta, \overline{\theta}, \theta \overline{\theta} , \theta^2, \overline{\theta}{}^2,  
\theta \overline{\theta}{}^2 ,\theta^2 \overline{\theta} ,\theta^2 \overline{\theta}{}^2 \}$.
Now the last $5$ columns have all entries equal to zero, except for one entry equal to $w_2$.
Similarly, the last $5$ rows have all entries equal to zero, except for one entry equal to $w_2$.
We calculate the determinant by expanding successively along each of the $5$ last columns, thereby
obtaining $5$ factors of $w_2$ and a sign (either plus or minus). 
With each expansion the corresponding row is also eliminated and this will eliminate the first $2$
of the last $5$ rows, but leave the remaining $3$ rows.
So we now expand along these remaining last $3$ rows, getting $3$ more factors of $w_2$ as well as a sign.
We then have remaining a $1 \times 1$ matrix whose entry comes from the 4th row and 4th column of the above matrix.
And that entry is again $w_2$. So the determinant of $G$ in the case $l=3$ is a sign times $9$ factors of $w_2$.
We leave it to the reader to check that the overall sign is positive and so we get $ \det G = (w_2)^9 = (w_2)^{3^2}$ as claimed.

Now we assume $l \ge 4$ and prove this case by induction.
The induction hypothesis that we will use is that in the case for $l-2$ the matrix $G$
(which is an $(l-2)^2 \times (l-2)^2$ matrix) has determinant $\pm (w_{l-3})^{(l-2)^2}$.
This is why we started this argument by proving separately the cases $l=2$ and $l=3$.

We start by using the same argument of expansion of the determinant as used above when $l=3$.
The matrix $G$ is an $l^2 \times l^2$ matrix. 
We consider this matrix in a basis made by ordering the basis $AW$
in such a way that the last elements are all of the form $\theta^{l-1} \overline{\theta}{}^k$
or of the form $\theta^{k} \overline{\theta}{}^{l-1}$, where $k \in I_l$.
Notice that the element $\theta^{l-1} \overline{\theta}{}^{l-1}$ is the only basis element in $AW$
that has both of these forms.
So, there are $2 \, \mathrm{card} (I_l) -1 = 2l -1$ such elements.
We claim that each of these elements has exactly one non-zero inner product with the elements in the ordered
basis $AW$ and that the value of that inner product is $w_{l-1}$.
Starting with $\theta^{l-1} \overline{\theta}{}^k$ we note that
$$
\langle \theta^i \overline{\theta}{}^j, \theta^{l-1} \overline{\theta}{}^k \rangle_w = \delta_{i+k, j+l-1} w_{j+l-1} \chi_l (j+l-1).
$$
This is zero if $j > 0$ because of the $\chi_l$ factor.
But for $j = 0$ we have
$$
\langle \theta^i , \theta^{l-1} \overline{\theta}{}^k \rangle_w = \delta_{i+k, l-1} w_{l-1}
$$
which is only non-zero for $i = l -1 - k$, in which case we have
$$
\langle \theta^{l-1-k} , \theta^{l-1} \overline{\theta}{}^k \rangle_w = w_{l-1}.
$$
A similar calculation shows that 
$$
\langle \overline{\theta}{}^{l-1-k} , \theta^{k} \overline{\theta}{}^{l-1} \rangle_w = w_{l-1},
$$
while all other elements in $AW$ have zero inner product with $ \theta^{k} \overline{\theta}{}^{l-1}$.
So we expand the determinant of $G$ along the last $2l-1$ columns, obtaining $2l-1$ factors of $w_{l-1}$ and some sign,
either plus or minus.
The corresponding rows that are eliminated, according to the above, are labelled by \textit{all} the powers of
$\theta$ alone or the powers of $\overline{\theta}$ alone.
Of these powers, only the two powers $\theta^{l-1}$ and $\overline{\theta}{}^{l-1}$
label one of the last $2l - 1$ rows. 
So we proceed by expanding along the remaining $2l -3$ last rows, thereby obtaining $2l-3$ more factors of  $w_{l-1}$
and some sign.
These $(2l -1) + (2l -3) = 4l -4 $ factors of $w_{l-1}$ as well as the sign multiply the determinant of a square matrix which has
$ l^2 - ( 4l -4  ) = (l-2)^2 $ rows and the same number of columns.

Now, we claim that we can calculate the determinant of this remaining $ (l-2)^2 \times  (l-2)^2 $ matrix,
call it $M$, using the induction hypothesis.
However, $M$ is not the matrix $G$ for the case $l-2$, but is related to it as we shall see.
From the labeling of the rows and columns of $G$ for the case $l$,
the matrix $M$ inherits a labeling, namely its rows and columns
are labeled by the basis elements $  \theta^i \overline{\theta}{}^j$ of $AW$ for $ 1 \le i,j \le l-2$.
This is clear by recalling the labeling of the columns and rows  which were eliminated in the above expansions.
But for any $a,b,c,d \in I_l$ we have that the entries of $G$ are
$$
\langle \theta^a \overline{\theta}{}^b , \theta^c \overline{\theta}{}^d \rangle_w = \delta_{a+d, b+c} w_{a+d} \chi_l (a+d).
$$
In particular, this holds for  $a,b,c,d \in I_l \setminus \{  0, l-1  \}$ in which case we can write
$$
\langle \theta^a \overline{\theta}{}^b , \theta^c \overline{\theta}{}^d \rangle_w = \delta_{a+d-2, b+c-2} w_{a+d} \chi_{l-2} (a+d-2),
$$
and so these are the entries in the matrix $M$.
By changing to new variables $a^\prime , b^\prime , c^\prime , d^\prime \in I_{l-2}$
where $a^\prime = a-1 , b^\prime = b-1, c^\prime = c-1 , d^\prime = d-1$ we see that the matrix entries of $M$ are
$$
 \delta_{a^\prime+d^\prime, b^\prime+c^\prime} w_{a^\prime+d^\prime +2} \chi_{l-2} (a^\prime+d^\prime).
$$
But the entries for the matrix $G$ in the case $l-2$ are
$$
  \langle \theta^{a^\prime} \overline{\theta}{}^{b^\prime} , \theta^{c^\prime} \overline{\theta}{}^{d^\prime} \rangle_w =
 \delta_{a^\prime +d^\prime , b^\prime +c^\prime } w_{a^\prime+d^\prime} \chi_{l-2} (a^\prime+d^\prime)
$$
for all $a^\prime , b^\prime , c^\prime , d^\prime \in I_{l-2}$.
So, except for a shift by $+2$ in the sub-indices of the weights,
the entries in $M$ correspond to the entries in $G$ for the case $l-2$.
Consequently, by the induction hypothesis as stated above, we have that 
$ \det M = \pm (w_{(l-3)+2})^{(l-2)^2} = \pm (w_{l-1})^{(l-2)^2}$.
Putting all this together, we have that 
$$
 \det G  = \pm (w_{l-1})^{(4l -4)} \det M  = \pm (w_{l-1})^{(4l -4)} (w_{l-1})^{(l-2)^2} = \pm (w_{l-1})^{l^2}
 $$
which proves our result. $\quad \blacksquare$

We leave it to the interested reader to track down the correct sign in the previous result.
We also  note that $w_{l-1}$ is the coefficient of $1$ in the definition of the weight
in (\ref{define-weight}), according to our convention.
Since  $w_{l-1} \ne 0$,
the main result of this section now follows immediately.
Here it is:

\begin{theorem}
For all $l \ge 2$ and for all $q \in \mathbb{R} \setminus \{ 0 \}$ the paragrassmann space $PG_{l,q}$ 
has a unique reproducing kernel with respect to the inner product
$\langle \cdot , \cdot \rangle_w$ defined in equation (\ref{rigor-define_form}).
\end{theorem}

\noindent
\textbf{Remark:}
$PG_{l,q}$ is a non-commutative algebra and so it is a quantum space
in the broadest interpretation of that terminology, that is,
it is not isomorphic to an algebra of functions defined on some set (`classical' space).
Also it is not even a Hilbert space with respect to the sesquilinear form that we are using on it.
Nonetheless, contrary to what one might expect from studying the theory of reproducing kernel
Hilbert spaces, this space \textit{does} have a reproducing kernel.

Also, we would like to comment that the existence of the reproducing kernel for  $PG_{l,q}$
is a consequence of the definition of the sesquilinear form and is not dependent on the
parameter $q$.

\section{Concluding Remarks}

In this paper we have introduced ideas from the analysis of reproducing kernel Hilbert spaces of functions
 to the study of the non-commutative space $PG_{l,q}$ of paragrassmann variables.
We are rather confident that other ideas from analysis will find application
to $PG_{l,q}$, and this will be one direction for future research.
Even more important could be the application of these ideas from analysis
to other classes of non-commutative spaces.
We expect that there could be many such applications.

Another possible direction for generalizing these results is to eliminate the nilpotency condition
and simply work in the infinite dimensional algebra
$\mathbb{C}  \{ \theta, \overline{\theta} \} / \langle \theta \overline{\theta} - q \overline{\theta} \theta \rangle$,
which is also known as the \textit{quantum plane}.
Greater care must be taken in this case and possibly some of our results will not hold
in complete generality.

The reproducing kernel for the Segal-Bargmann space $\mathcal{B}_H$  will be used in a subsequent paper \cite{part2} to
define Toeplitz operators in that space with symbols in the space $PG_{q,l}$. 

\vskip 0.45cm \noindent
\textbf{\Large Acknowledgments}
\vskip 0.35cm
This paper was inspired by a talk based on \cite{csq} given by Jean-Pierre Gazeau during my sabbatical stay
at the Laboratoire APC,  Universit\'e Paris Diderot (Paris 7) in the spring of 2011. 
Jean-Pierre was my academic host for that stay, and so I thank him not only for stirring my curiosity in this subject
but also for his most kind hospitality which was, as the saying goes, above and beyond the call of duty. 
(I won't go into the details, but it really was. Way beyond, actually.)
Merci beaucoup, Jean-Pierre!
Also my thanks go to Rodrigo Fresneda for very useful comments
as well as for being my most gracious host at the UFABC in S\~ao Paulo, Brazil in April, 2012
where work on this paper continued.
Muito obrigado, Rodrigo!

\end{document}